\documentstyle[12pt,aasms4]{article}
\begin{document}
\received{:}
\accepted{:}

\lefthead{Vilhu and Nevalainen}
\righthead{
    Rings of GRS 1915+105}

\title{Two-phase modelling of the rings in the RXTE two-color diagram of 
            GRS 1915+105}

\author{Osmi Vilhu}
\affil{ Observatory, Box 14, FIN-00014 University of Helsinki, Finland}
\and
\author{ Jukka Nevalainen}
\affil{Harvard-Smithsonian Center for Astrophysics, 60 Garden Street, 
                  Cambridge, MA 01238, USA,  and Observatory, University of 
                  Helsinki, Finland}

\begin{abstract}

The Galactic superluminal source GRS 1915+105 was found to experience
a peculiar X-ray variability in a narrow count rate range (9300 - 12100 cts/s,
 5 PCU's)
 of the Proportional Counter Array (PCA) onboard the 
Rossi X-ray Timing Explorer (RXTE).
This can be seen as a ring-shaped pattern in the two-color diagram of count
rates, where the
hard hardness F(13-40keV)/F(2-13keV) is plotted against the soft hardness
F(5-13keV)/F(2-5keV). The system runs one cycle with  periods
ranging between 50 - 100 s for different observations, one rotation 
in the 2-color diagram corresponding 
to the time between two contiguous maxima in the light curve.     
We model this behaviour 
successfully
with the help of a self-consistent 2-phase 
thermal model where  seed photons from an optically thick classical 
disk are Comptonized in a hot spherical corona surrounding the inner disk   
( \cite{pou96,vil97,nev98}).  
In the model, changes of two parameters regulate the paths in the 
2-color diagram: the black body temperature T$_{in}$ of the inner disk and 
the Thomson optical depth multiplied by the electron temperature  
of the hot phase $\tau$T$_e$. These parameters oscillate with time 
but with a  phase-shift between each other, causing the    
ring-shaped pattern. 
During  the observation studied in  more detail (20402-01-30-00), 
the inner disk radius  varied with 97 s period 
between 20 - 35 km with an anticorrelation
between the coronal $\tau$T$_e$ and the mass accretion rate $\dot M$
through the disk,
possibly indicating a coupling between the disk and coronal accretion. 
During a typical cycle, the inner disk radius rapidly shrinked 
and returned more slowly back to the original larger value.
In the rings we may see phenomena close to the black hole horizon under near
Eddington accretion rates.


\end{abstract}

\keywords{accretion, accretion disks---binaries:close---black hole physics---
          instabilities---X-rays:stars---stars:individual:GRS 1915+105}

\section{Introduction}

The X-ray transient GRS 1915+105 was discovered by \cite{cas92}. 
\cite{mir94} found superluminal jets at an angle of 70$^0$ to the line of sight
in this  Galactic black hole candidate which lies at a kinematic
distance of 12.5 $\pm{1.5}$ kpc (\cite{cha96}). 
The Rossi X-ray Timing Explorer has been monitoring it frequently, making the
data openly available to the scientific community.
In this way
GRS 1915+105 has become an important target for studying accretion phenomena
close to a potential black hole. 

A rich pattern of variability has already emerged from the RXTE data
(\cite{gre96,mo97,che97,taa97,bel97a,bel97b,mar97}).
In particular, \cite{bel97b} interpreted
the variability during their observations (mean 5 PCU count rate 14000 cts/s) 
as a rapid 
disappearance  of the inner disk  (through the horizon), 
and suggested that the viscous time scale controls the variability
in this radiation-pressure dominated region .
They modelled the  spectra in the 2 - 40 keV range 
with the classical multi-temperature disk  with a
power-law component. This modelling allowed the study of the changes of
the inner disk radius which, during the 
observations, varied between 20 - 80 km.

In the present letter we report a variability of GRS 1915+105
which manifests itself as a ring-shape pattern in the  2-color
diagram. Further, we model this behavior with the
help of a self-consistent two-phase model where seed photons from a 
classical disk are Comptonized in a spherical  corona surrounding the inner
disk. 

\section{Observations}
We collected several Proportional Counter Array (PCA) observations from the 
TOO-archive of RXTE, 
with typical durations of a few hours. 
The Standard 2 data with 128 channels of spectral information
     and 16-sec temporal resolution were used. 
     The background was subtracted although
     its effect was not crucial (being around 10 per cent in the HR2-color).
The data were binned into 16 sec time-bins and further
(following \cite{bel97b}) into
two X-ray colors HR1 = B/A and HR2 = C/(A + B), where A, B and C are  
the counts in the 2.0 - 5.0 keV (channels 3-10), 5.0 - 13.0 keV 
(channels 11-32) and 13.0 - 40.0 keV (channels 33-77) 
intervals, respectively. Color-color diagrams were  constructed 
showing many different patterns with a strong dependence on the average 
count-rate. 

Particularly interesting features were the {\bf rings}, 
occurring
at the rising phase of the RXTE/ASM light curve (see Fig.1 and Table 1) and
having mean PCA count rates in a narrow interval 9300 - 12100 cts/s 
(2 - 40 keV, 5 PCU's).   
These are discussed in the next Chapter. 
 They form a sub-class of the quasi-periodic structure in the time domain,
    discussed earlier in the references cited above and showing hysteresis
    behavior and closed loops in the 2-color diagram. We call them as 'rings',
    separating them from the other observations, due to their rather regular 
    and ring-shaped paths in the 2-color diagram. For comparison,
    the 2-color diagram of the observation analysed by \cite{bel97b} 
    looks like a crescent.

\placetable{tbl-1}
\placefigure{fig1}

\section{Rings in the X-ray 2-color diagram and their modelling}

Table 1 and Figure 2 summarize the seven observations showing ring-behavior. 
For comparison, the observation analyzed by \cite{bel97b},
and one observation in the long quiescent lull-phase, are indicated
(see the dotted lines in Fig.1). 
\placefigure{fig2}
\placefigure{fig3}
\placefigure{fig4}
 The system runs clock-wise in the 2-color diagram with average periods
   between 47 - 97 s (see Table 1), computed as the total observing time 
   divided by the number of maxima in the light curve. 
   This number of maxima equals to
   the number of rotations in the 2-color diagram, since one rotation 
   corresponds to the time between two contiguous maxima. Timing work 
   with better time resolution is in progress, in the present letter
   we concentrate on average (over 16 s) behavior of the oscillations.

Figures 3 and 4 give  more details of one particular observation
(20402-01-30-00) and its time evolution. This 
observation was selected due to its long period (97 sec), allowing to construct
reasonably accurate light curves despite 
the 16-s time resolution.
 The fastest speed (angular velocity) along the ring of this particular
observation
was on the average 80 degrees/16 sec (as seen from the center of the ring). 
The 16 sec averaging should be statistically adequate to characterise 
the ring when many data-points from different phases are collected. 
 
We attempted to model the rings for this letter. A more detailed spectral
fitting will be presented in a forthcoming paper, including  
simultaneous HEXTE and time-averaged BATSE data. 
The simplest way would be
to use the multi-color disk black body model with a power-law component, as
done in some of the papers cited above. However, the power-law component alone 
includes no physics nor system geometry, although a popular explanation 
for its origin is the  Comptonization process.
 
Instead, we used a two-phase thermal radiative model to couple the disk 
and Comptonized radiation. In the model, the cool phase consists of
an optically thick multi-temperature  disk, generating
the observed soft spectrum and serving as a source of the seed photons
Comptonized in the spherical 
hot phase (corona) surrounding the inner disk
(giving rise to the power-law component). The geometry resembles and 
approximates
that of the advective disks as developed e.g. in \cite{esi97}. The radiative
transfer was handled by the iterative scattering method (ISM) developed by
\cite{pou96} and incorporated into the XSPEC spectral fitting software
of the XANADU package (for more details of the model, see \cite{vil97,nev98}).
 In a specific geometry (a fixed ratio of the inner disk and
coronal radii, = 0.5 used here) 
the model is self-consistent and predicts the normalization
between the soft and power-law components, as well as the power-law photon 
index.

The free model parameters are the black body temperature T$_{in}$
at the inner disk  (with the temperature-stratification in the disk as
T(r) = T$_{in}$(r/R$_{in}$)$^{-3/4}$), 
the electron temperature T$_e$ and
the Thomson optical depth $\tau$ of the hot corona.
 The inclination  was fixed to
70$^0$ (\cite{mir94}), assuming that the jets are perpendicular to the disk. 
Information of the high energy cut-off
is needed to fix the coronal temperature which is  possibly 
beyond the RXTE capability (in a forthcoming paper we try to fix it with the
HEXTE data).
In the narrow PCA energy interval 2-50 keV, $\tau$ and T$_e$ cannot be separated
but their product $\tau$T$_{e}$ uniquely specifies the spectrum together 
with T$_{in}$.
The Hydrogen column was fixed at N$_H$ = 4.5 10$^{22}$ cm$^{-2}$ as in 
\cite{bel97b}. 
This value was adopted since it is close to our preliminary fits.
   By changing the value, one shifts the model curves of Fig.3 in the North-East -
  South-West direction. Using N$_H$ = 3.5 10$^{22}$ cm$^{-2}$
  gives 1.4 and 1.2 times larger T$_{in}$ and $\tau$ values, respectively,
  having no qualitative effect on our conclusions. Just the parameter
  values would be somewhat different. 
 
The curves added to Fig.3 show the above reference model
as a function of T$_{in}$ and $\tau$$_{50}$ = $\tau$T$_e$/50keV. 
The effects of these two 
parameters act in the 2-color diagram 
perpendicularly to each other. The cycling behavior can be 
understood as a variability of these two  parameters with a phase shift 
(the maxima
or minima do not coincide in time). 
The model gives the disk luminosity (proportional to
 R$_{in}$$^2$T$_{in}$$^4$) from 
which the inner disk radius R$_{in}$ can be computed
(assuming d = 12.5 kpc). 
The light curves of the observation 20402-01-30-00  
are shown in Fig.4, as interpolated from the reference model curves in Fig.3.

\section{Discussion}

          We have analysed the data using the 16 s time resolution
      of Standard 2 data. Hence, any existing sub-structure (like
      in the observation by \cite{taa97} was smoothed
      out. However, the time resolution used is sufficient to demonstrate
      the qualitative nature and shape  our ring-data in the 2-color diagram 
      (see Figures 2 and 3) and the applicability of our  
      modelling at this resolution. Further work is clearly needed for a more
      detailed timing and spectral analysis of the observations in Table 1.

The changes of  T$_{in}$ and $\tau$$_{50}$ = $\tau$T$_e$/50keV in our modelling 
are  sufficient to explain
the whole pattern of different paths in the 2-color diagram.    
The ratio of the soft and hard luminosities, the power-law
photon index $\alpha$ and $\tau$$_{50}$ are  strongly correlated
(for small values of $\tau$ and T$_e$, the Kompaneets
y-parameter is close to their product). 
The spectrum in the PCA energy-interval 
is determined by  T$_{in}$ and $\tau$$_{50}$, and 
$\alpha$ is uniquely determined by $\tau$$_{50}$ 
(e.g. $\alpha$ = 3.2 and 2.5 when
$\tau$$_{50}$ = 0.5 and 1.0, respectively). 
To check the consistency we run the observations analyzed
by \cite{bel97a} through our models in the 2-color diagram.
This resulted in a qualitatively similar photon index behavior
as a function of time to that which was obtained with the
diskbb + power-law modelling. 
Preliminary studies of
individual ring-spectra (accumulated
by count rate criteria) also confirm our modelling but these will be 
the topics of a separate paper.  

The variability is probably controlled by the viscous time scale
as demonstrated for one set of observations by \cite{bel97b},
 although it remains to be studied whether this is the case
for all observations, in particular for the rings discussed in the
present paper.

The mass accretion rate (through the disk) can be computed from the formula of 
classical viscous disks (\cite{fra92}):
$\dot M$ = 8$\pi$R$_{in}$$^3$$\sigma$T$_{in}$$^4$/3GM.
The rates computed from this formula are high and 
the inner disk lies 
inside the unstable radiation-dominated region of the thermal balance curve.
 Because of this, the disk is probably advection dominated
and the mass transfer rates computed  may be in error.
Further, at very high rates the accretion through the disk and corona
may anticorrelate (\cite{esi97}, their Sec.3.4) and the true accretion rate
might behave quite differently from that computed from the above formula. 

To obtain a better idea of what is happening, mean light curves of important 
parameters were accumulated along the ring-path of Fig.3 (see Fig.5). 
\placefigure{fig5} 
The whole ring-cycle can be visualized in the following way 
(see Fig.5):
 
Locally the inner disk is unstable, perhaps because it is  inside
the radiation pressure dominated zone.
The mass transfer rate $\dot M$ starts to increase from its value 
0.15$\dot M_{Edd}$ (of 10M$_{\odot}$ using the classical formula above and
$\dot M_{Edd}$ = L$_{Edd}$/c$^2$ ) 
leading to slow
increases of the luminosity and  the inner disk radius.  
At a certain point the inner disk  starts to 
shrink rapidly, leading to the count rate spike and a clear minimum of the 
coronal $\tau$$_{50}$. At the same time  
$\dot M$ reaches its maximum 0.5$\dot M_{Edd}$ and 
returns rapidly  to its smaller original value.
After the minimum radius has been reached the
mass transfer rate starts to increase again and a new cycle starts.

$\tau$$_{50}$ and $\dot M$ were determined quite independently, so their
anticorrelation in Fig.5 is remarkable. This might indicate that a fraction
of the mass accretion is channelled through the disk, and another fraction
through the corona (reflecting its optical depth $\tau$$_{50}$). 
During the whole process
the total accretion may remain constant, just its different contributions 
vary. 

We note that during each individual cycle, the system spends half of
its time in the lower-right corner of Fig. 3 (a state of minimum count rate),
and only rapidly visits the
state of high count rates. The rising times are also longer
than the decay times. 
Sometimes, but not always, the system seems to spend additional time during
count rate and R$_{in}$ maxima. 
This time-asymmetry may give some additional
information for the reason of the peculiar behavior of the system.
The picture of a rapid disappearance of the whole inner disk annulus 
through the
horizon (as in the \cite{bel97b} case) is not so simple here, since
the emptying process (the time evolution of increasing R$_{in}$) is 
slow (see Fig.5). Of course, after the rapid shrinking phase, some part of the
inner disk could have disappeared 
(note in Fig.5, the perhaps not so significant short rise after the minimum).  

The main difference between  the rings discussed here and the 
observations analyzed
by \cite{bel97b} is that in the latter case the amplitudes of 
T$_{in}$ and $\tau$$_{50}$ are in phase (the  minima  of $\tau$$_{50}$
correspond to the maxima of the count rate and T$_{in}$). For the rings,
instead, there is a phase shift between these two parameters. 
This follows simply by converting (at each time bin) the observed 
          pairs of (HR1,HR2)-values into the corresponding pairs of (T$_{in}$,
          $\tau$$_{50}$)-values, using the theoretical grid shown in Fig.3. 
          In the \cite{bel97b} observations, T$_{in}$ maxima and
          $\tau$$_{50}$ minima coincide (with no phase-shifts), following from
          the  crescent-like form of the 2-color diagram 
          (see the dots in Fig.2). Future
          work on disk instabilities should explain the physical reasons
          of these different paths. 
          
Further, the mean count rate, the mean inner disk radius and its amplitude
were larger in the \cite{bel97b} case. 
 Hence, the ring-behavior might
be characteristic to disk instabilities very close to the innermost stable
Keplerian orbit,     where the relativistic potential barrier is small 
and just 
a small change in angular momentum leads to accretion through the horizon
(\cite{fra92}).
 However, 
the minimum values of R$_{in}$ obtained (20 km) are too small for a 
non-rotating black hole and 
require a rotating Kerr hole to be related to the innermost stable 
orbit, 
unless the central mass is smaller than 2-3 solar masses.

\section{Conclusions}
We found a peculiar form of variability ({\bf rings}) of GRS 1915+105 in
the RXTE/PCA 2-color diagram (see Table 1 and the Figures 2 and 3). This
pattern was observed in a narrow interval 9300 - 12100 cts/s (2 - 40 keV,
5 PCU's),
intermediate between the very high and lull states (see Fig.1). The mean
period of variability is between 50 - 100 sec. There is no direct correlation
between the cycle-period and count rate, although a real variability
(spin-ups and spin-downs) may have taken place. The system spent most of the
time in the low count rate level and only rapidly visited the state of high
count rates (with longer rise than decay times).   

We modelled this behavior with a 2-phase self-consistent radiative thermal 
model where an optically thick multi-temperature disk gives rise to the
soft part of spectrum and serves as a source of  seed photons, Comptonized
in the hot spherical corona surrounding the inner disk. We explain the 
variability with the help of two parameters (T$_{in}$ and $\tau$$_{50}$ =
$\tau$T$_e$/50keV), and
with a phase-shift between their light curve amplitudes. 

During  the observation studied in  more detail (20402-01-30-00), 
the inner disk radius R$_{in}$ oscillated between 20 - 35 km,
 the smallest value was reached somewhat after the count rate maximum and 
$\tau$$_{50}$-minimum. 
During a typical cycle, the inner disk radius shrinked, 
after which returned more slowly back to the original larger value (see Fig.5).
$\dot M$ and $\tau$$_{50}$ anticorrelated, possibly
indicating a balance between the accretion flows through the disk and corona
(in the sense that the total accretion rate might remain constant).
In the rings we may see interesting 
phenomena very close to the black hole horizon during high accretion rates.

\acknowledgements

We are grateful to the whole staff of the RXTE Guest Observer Facility
at GSFC, especially to Tess Jaffe, Gail Rohrbach and Padi Boyd, for the
great and friendly assistance during our stay at GOF. We thank  Craig
Markwardt for discussions of his interesting results, Juri Poutanen 
for his ISM models and most valuable help and advice during their usage and
interpretation of the results, and Diana Hannikainen for reading the manuscript
and for comments. We also thank the referee for useful critisism.  
JN thanks the Harvard Smithsonian Center for Astrophysics for hospitality and
the Smithsonian Institute for a Predoctoral Fellowship and the Academy 
of Finland for a supplementary grant.

\clearpage
 
\begin{deluxetable}{crrrrrrrrrrr}
\footnotesize
\tablecaption{RXTE/PCA ring-data. \label{tbl-1}}
\tablewidth{0pt}
\tablehead{
\colhead{ID} & \colhead{JD - 2450000}  
            & \colhead{5 PCU cts/s}   & \colhead{cycle time  (sec)}   
} 
\startdata
20402-01-30-00 &595.035 &  9260 &  97 \nl
20402-01-31-00 &603.083 & 10300 &  74 \nl
20402-01-31-01 &604.986 & 10420 &  49 \nl
20402-01-31-02 &605.135 & 10540 &  47 \nl
20402-01-32-01 &608.847 & 11320 &  67 \nl
20402-01-32-00 &608.985 & 11300 &  82 \nl
20402-01-34-01 &622.333 & 12150 &  53 \nl
\enddata
\end{deluxetable}

%
%

\clearpage

\figcaption[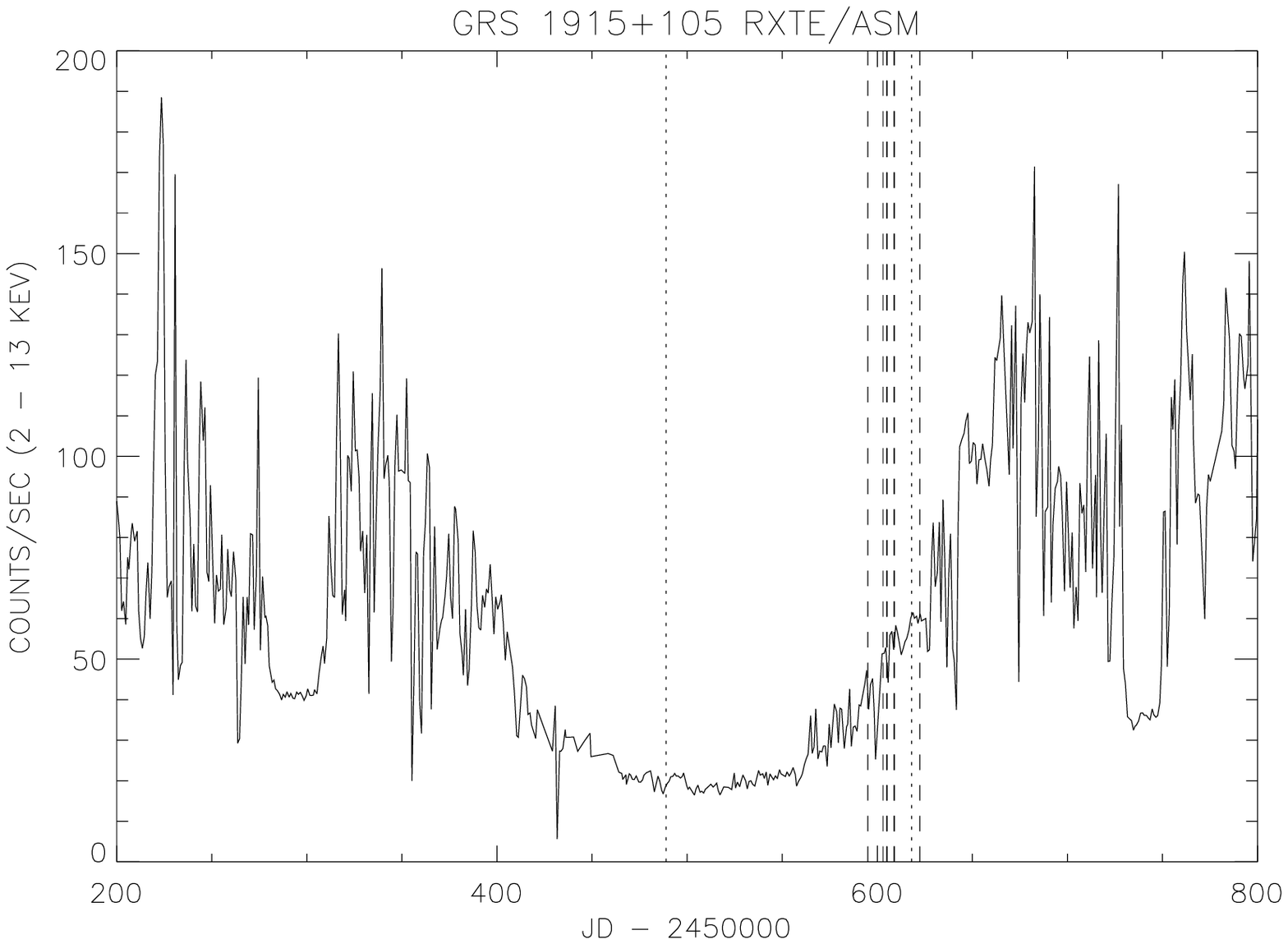]{The ASM light curve (2 - 13 keV) of GRS 1915+105
     with the ring-observations of Table 1 marked with vertical dashed lines.
The dotted lines show the times of other observations discussed in the text
(and marked in Fig.2).
                              \label{fig1}}
\figcaption[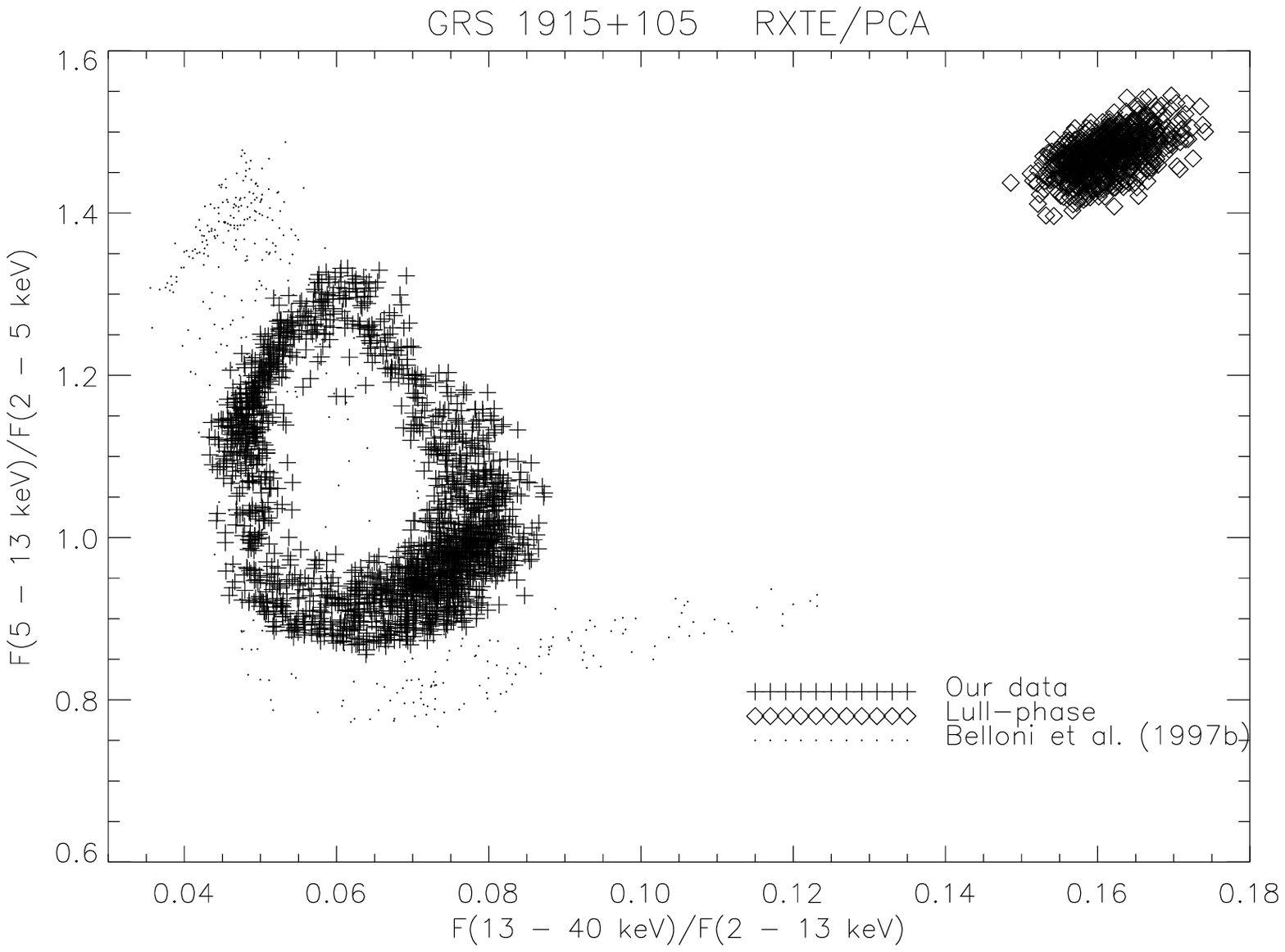]{The ring-data of Table 1 plotted 
        in the 2-color diagram with 16 s time-binning (plus-signs).
   For comparison, the observation analyzed by \cite{bel97b} (marked
with points) and one observation
   from the long lull-phase of Fig.1 (the small patch at the upper-right marked with diamonds) 
       are overplotted.   \label{fig2}}

\figcaption[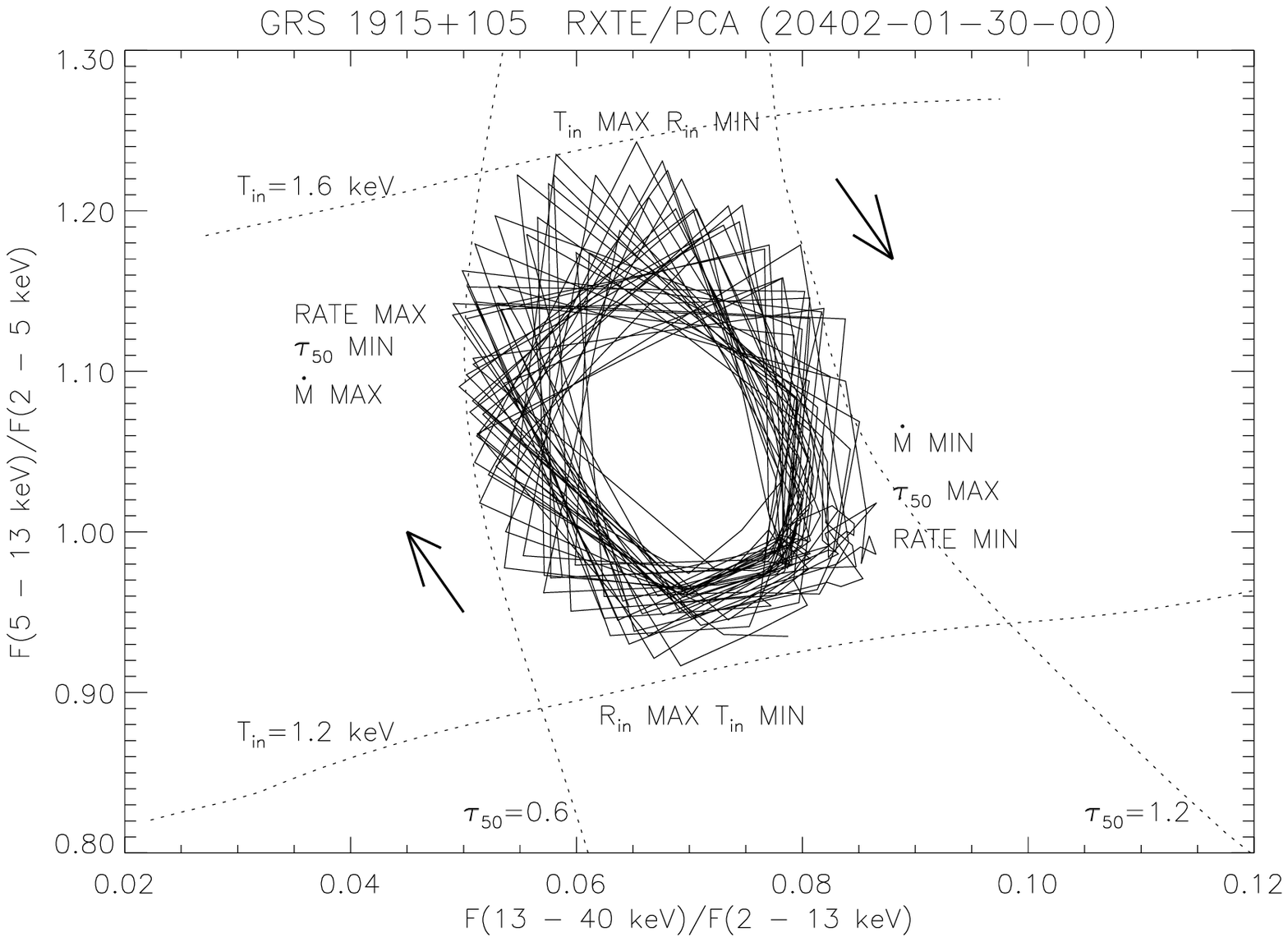]{ The observation 20402-01-30-00 
  (see Table 1) 
  plotted in the 2-color diagram with 16 sec time-binning 
(neighboring data-points in the light curve
joined with lines). The big arrows show
 the clock-wise evolution with 97 s mean cycle period.    
 Positions of maximum and minimum values of some variables are marked.
 The 2-phase model curves (see the text) are overplotted with dotted lines.
       \label{fig3}}

\figcaption[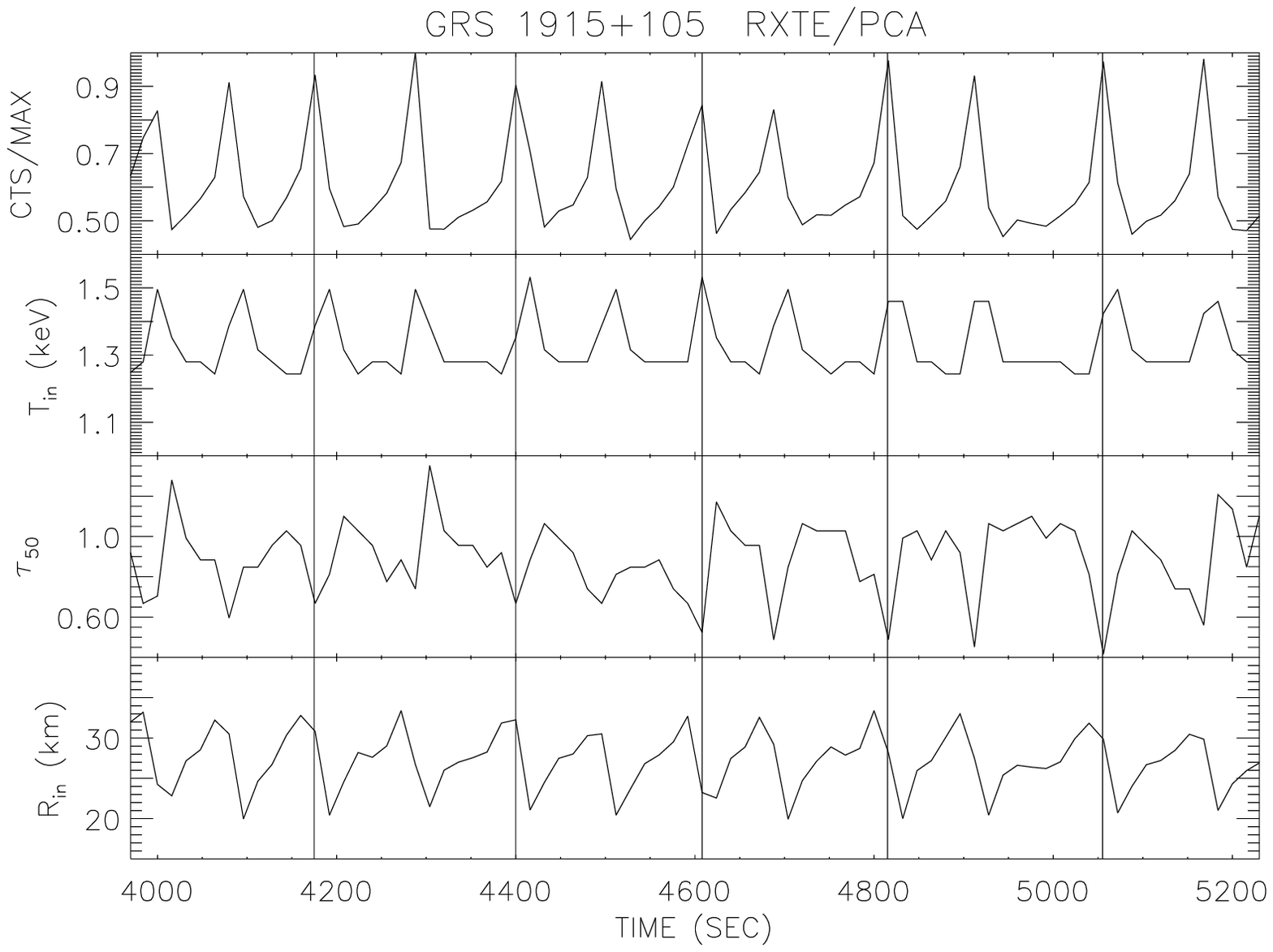]{Selected light curves for the observation
20402-01-30-00 (from Fig.3):
the total count rate 2 - 40 keV (5 PCU's, in units of 15000 cts/s), 
the inner disk temperature
 T$_{in}$, the Thomson optical depth
  $\tau$ multiplied by the electron temperature T$_e$ of the spherical
corona ($\tau$T$_e$/50keV) and the inner disk radius R$_{in}$.    \label{fig4}}

\figcaption[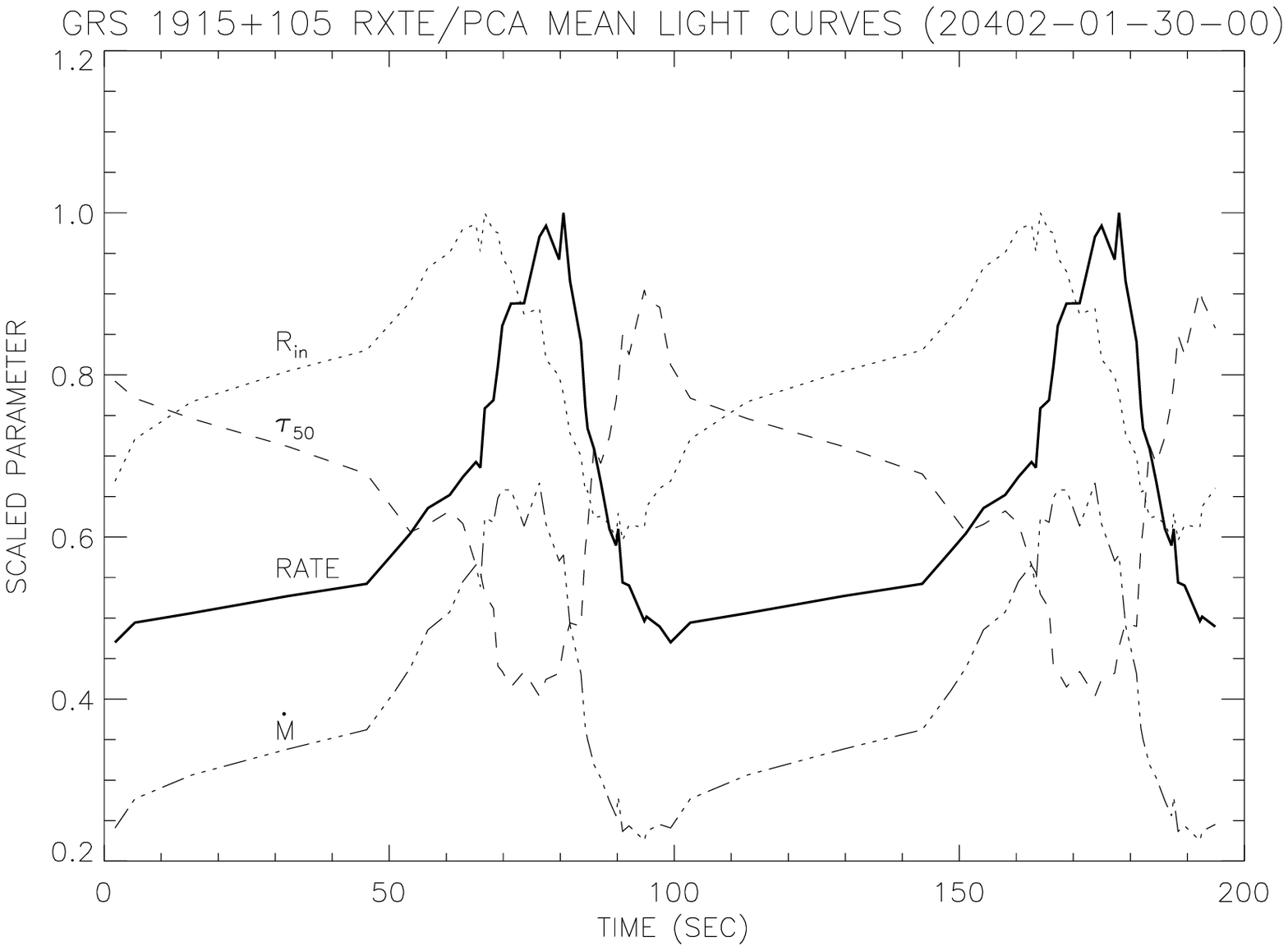]{Mean light curves over one cycle for the observation
of Fig.3. The parameters are scaled by the following values (in parentheses):
RATE (15000 cts/s (5 PCU units), solid line), R$_{in}$ (35 km, dotted line), 
$\tau$$_{50}$ (1.4, dashed line) and
$\dot M$ (0.75L$_{Edd}$/c$^2$ of 10M$_{\odot}$, dash-dotted line). 
Two identical cycles are plotted 
for clarity. \label{fig5}}




\end{document}